\documentstyle{elsart}

\input{epsf}

\newcommand{\ABO}{A $+$ B $\to$ 0}
\newcommand{\ABOS}{A $+$ B(static) $\to$ 0}

\newcommand{\EA}{\eta_{\mathrm{A}}}
\newcommand{\EB}{\eta_{\mathrm{B}}}
\newcommand{\ER}{\eta_{\mathrm{R}}}
\newcommand{\SA}{S_{\mathrm{A}}}
\newcommand{\SB}{S_{\mathrm{B}}}
\newcommand{\SR}{S_{\mathrm{R}}}
\newcommand{\DA}{D_{\mathrm{A}}}
\newcommand{\DB}{D_{\mathrm{B}}}
\newcommand{\JA}{J_{\mathrm{A}}}
\newcommand{\JB}{J_{\mathrm{B}}}
\newcommand{\GA}{\gamma_{\mathrm{A}}}
\newcommand{\GB}{\gamma_{\mathrm{B}}}
\newcommand{\RA}{\rho_{\mathrm{A}}}
\newcommand{\RB}{\rho_{\mathrm{B}}}
\newcommand{\CA}{C_{\mathrm{A}}}
\newcommand{\CB}{C_{\mathrm{B}}}
\newcommand{\Cf}{C_{\mathrm{f}}}
\newcommand{\Cw}{C_{\mathrm{w}}}
\newcommand{\CJ}{C_{\mathrm{J}}}
\newcommand{\dc}{d_{\mathrm{c}}}
\newcommand{\xf}{x_{\mathrm{f}}}

\newcommand{\W}{\tilde{w}}
\newcommand{\WW}{\W_1}

\newcommand{\PXX}[1]{
   \frac{\textstyle\partial^2{#1}}{\textstyle\partial x^2}}
\newcommand{\PT}[1]{
   \frac{\textstyle\partial{#1}}{\textstyle\partial t}}

\newcommand{\BRA}[1]{\!\left( #1 \right)}
\newcommand{\SQR}[1]{\!\left[ #1 \right]}

\newcommand{\erf}[1]{ \:\!\mbox{erf}\!\left(#1\right)}

\newcommand{\ierfc}[1]{\:\!\mbox{ierfc}\!\left(#1\right)}
\newcommand{\ierfs}[1]{\:\!\mbox{erf$^{-1}$}\left[#1\right]}

\begin{document}


\begin{frontmatter}
 \title{
         The asymptotic behaviour \\
         of the initially separated \\
         \ABOS\ reaction-diffusion systems
       }
  \author{Zbigniew Koza\thanksref{KOZA}}
  \address{Institute of Theoretical Physics,
             University of Wroc\l{}aw,\\
            50-204 Wroc\l{}aw, Poland.
          }
  \thanks[KOZA]{e-mail: zkoza@ift.uni.wroc.pl}

\begin{abstract}
   We examine the long-time behaviour of \ABOS\ reaction-diffusion
   systems with initially separated species A and B. All of our analysis
   is carried out for arbitrary (positive) values of the diffusion
   constant  $\DA$ of particles A and initial concentrations $a_0$ and
   $b_0$ of A's and B's. We derive general formulae for the location of
   the reaction zone centre, the total reaction rate, and the
   concentration profile of species A outside the reaction zone. The
   general properties of the reaction zone are studied with a help of
   the scaling ansatz. Using the mean-field approximation we find the
   functional forms of `tails' of the reaction rate $R$ and the
   dependence of the width of the reaction zone on the external
   parameters of the system. We also study the change in the kinetics of
   the system with $\DB>0$ in the limit $\DB\to0$. Our results are
   supported by numerical solutions of the mean-field reaction-diffusion
   equation.
\end{abstract}


\end{frontmatter}




\section{Introduction}

Investigation of the reaction fronts formed in the reaction-diffusion
processes of the type \ABO\ with initially separated species A and B has
attracted a lot of recent interest. This is due to the fact that not
only the kinetics of such systems exhibits many surprising features
\cite{G-R,Overview,HaimExperiment,Hav95,Koza96}, but they are also
amenable to experimental studies
\cite{HaimExperiment,Experiment,HaimKudowa}.

A standard way to treat the initially separated problem analytically is
to solve the partial differential equations \cite{G-R}
\begin{equation}
  \label{GR}
    \begin{array}{rcl}
      \PT{\RA} &=& \DA \PXX{\RA} - R \,, \\[2ex]
      \PT{\RB} &=& \DB \PXX{\RB} - R \,,
     \end{array}
\end{equation}
with the initial state given by
\begin{equation}
   \label{IniCond}
     \begin{array}{rcl}
        \RA(x,t=0) &=& a_0 H(-x) \,,\\[1ex]
        \RB(x,t=0) &=& b_0 H(x)  \,,
     \end{array}
   \end{equation}
where $\RA(x,t)$ and $\RB(x,t)$ are the mean local concentrations of A's
and B's, $R$ is the macroscopic reaction rate, $H(x)$ denotes the
Heavyside step function, and $a_0$, $b_0$, $\DA$ and $\DB$ are some
constants related to the initial concentrations of species A and B and
their diffusion coefficients, respectively.

Equations (\ref{GR}) present the macroscopic approach to the problem.
However, it has not been established yet how to relate the macroscopic
reaction rate $R$ to the microscopic picture of the initially separated
reaction-diffusion systems below or at the critical dimension $\dc=2$.
Dimensional \cite{CD-Steady,Krapivsky,CDC91} and renormalisation group
analyses \cite{RG-Rapid,RG-Front,RG-Homo} lead to the conclusion that
above $\dc$ one can adopt the mean-field approximation $R=k\RA\RB$, with
$k$ being a constant. For 2D systems one expects logarithmic corrections
to the mean-field picture \cite{Krapivsky,CDC91,RG-Homo}. One
dimensional systems are usually studied by examining microscopic models
in which, upon contact, the members of a pair A--B react with some
probability $p$
\cite{CDC91,RG-Rapid,RG-Front,RG-Homo,Bstatic,Cornell95,LAHS3-8,%
Barkema,Argentina}. The analytical form of $R(x,t)$ was derived for $\DB
= 0$ \cite{Bstatic}, and for $\DB > 0$ and $|x| \to \infty$
\cite{RG-Front}.

There are, however, several techniques which enables one to derive a lot
of useful information from (\ref{GR}) even for $d\le \dc$, i.e. when the
explicit form of $R$ remains unknown. They are concentrated on the
asymptotic, long-time behaviour of the reaction-diffusion systems, and
include the renormalisation group
analysis \cite{RG-Rapid,RG-Front,RG-Homo,Barkema}, the scaling ansatz
\cite{G-R,CDC91}, the quasistationary approximation
\cite{CD-Steady,BenRedner}, and our approach of Ref.~\cite{Koza}.

According to the scaling ansatz \cite{G-R}, the long-time behaviour of
the reaction-diffusion system inside the reaction layer can be described
with a help of some scaling functions $\SA$, $\SB$ and $\SR$ through
\begin{eqnarray}
  \label{SA}
    \RA(x,t) &=&\EA
          t^{-\GA}\SA\BRA{x - \xf(t) \over w(t)}\,,\\[1ex]
  \label{SB}
    \RB(x,t) &=& \EB
         t^{-\GB}\SB\BRA{x - \xf(t) \over w(t)}\,,\\[1ex]
  \label{SR}
      R(x,t) &=& \ER
         t^{-\beta}  \SR\BRA{x - \xf(t) \over w(t)}\,,
\end{eqnarray}
where $\xf(t)\propto t^{1/2}$ denotes the point at which the reaction
rate $R$ attains its maximal value, $w(t) \propto t^\alpha \ll t^{1/2}$
is the width of the reaction zone, $\EA$, $\EB$ and $\ER$ are some
parameters independent of $x$ and $t$, and exponents $\alpha$, $\beta$,
$\GA$ and $\GB$ are some positive constants given, for $R
\propto \RA\RB$ and nonzero diffusion constants, by $\alpha =
\frac{1}{6}$, $\beta = \frac{2}{3}$ and $\GA = \GB =
\frac{1}{3}$.

The quasistationary approximation \cite{CD-Steady,BenRedner} consists in
the assumption that at sufficiently long times the kinetics of the front
is governed by two characteristic time scales. One of them,
$\tau_{\mathrm{J}} \propto (\!{\d}\log\! J/ \!{\d} t)^{-1} \propto t$,
controls the rate of change in the diffusive current $J$ of particles
arriving at the reaction layer. The other one, $\tau_{\mathrm{F}}
\propto w^2 \propto t^{2\alpha}$, is the equilibration time of the
reaction front. If $\alpha < 1/2$ then
$\tau_{\mathrm{F}}/\tau_{\mathrm{J}} \to 0$ as $t \to \infty$.
Therefore, as $t\to\infty$, in the vicinity of $\xf$ the left sides of
(\ref{GR}) become negligibly small compared to other terms.
Consequently, if $\DA$ and $\DB$ are both nonzero, the asymptotic form
of $\RA$ and $\RB$ inside the reaction layer is governed by much simpler
equations
\begin{equation}
  \label{STAC}
     \begin{array}{rcl}
       \DA \PXX{\RA} &=& R \,,\\[2ex]
       \DB \PXX{\RB} &=& R \,,
     \end{array}
\end{equation}
which are to be solved with the boundary conditions
\begin{equation}
  \label{BC}
    \begin{array}{rcl}
      \partial\RA/\partial x \to -\JA(t),\; \RB \to 0   &\; \mbox{as}\;&
         x \to  -\infty\,,\\[1ex]
      \partial\RB/\partial x \to \JB(t),\;\RA \to 0     &\;\mbox{as}\;&
         x \to +\infty\,.
    \end{array}
\end{equation}
The most important feature of the quasistationary equations (\ref{STAC})
is that they depend only on $x$, with time $t$ being a parameter
entering their solutions $\RA(x,t)$ and $\RB(x,t)$ only through the time
dependent boundary currents $\JA = \JB$ whose dependence on $t$, $\DA$,
$\DB$, $a_0$ and $b_0$ has recently been derived analytically
\cite{Koza}.

In a recent paper \cite{Koza} we employed the quasistatic approximation
to develop a new method of investigating the asymptotic
kinetics of the initially separated reaction-diffusion systems. A
peculiar feature of that approach is that it is concentrated mainly on
the properties of the system outside the reaction zone. In this way,
without imposing any special restrictions on the form of $R$, it relates
{\em exactly\/} many quantities of physical interest to the values of
external parameters $a_0$, $b_0$, $\DA$ and $\DB$, which will enable us
to investigate the limit $\DB\to 0$ analytically. In particular it was
shown that there exist two limits, $\Cf = \lim_{t\to\infty}
\xf(t)/\sqrt{t}$ and $\CJ = \lim_{t\to\infty} J(t)\sqrt{t}$. Given
$a_0$, $b_0$, $\DA$ and $\DB$, the value of $\Cf$ can by computed by
solving
\begin{equation}
  \label{cfeq}
   \Phi\!\left(
             \frac{-\Cf}{2\sqrt{\DA}}
       \right)
      =
        \frac{a_0\sqrt{\DA}}{b_0\sqrt{\DB}} \:
        \Phi\!\left(
                  \frac{\Cf}{2\sqrt{\DB}}
            \right) \,,
\end{equation}
where
\begin{equation}
  \label{defphi}
  \Phi (x) \equiv \SQR{1 - \erf{x}}\exp(x^{2})\,,
\end{equation}
and $\;\erf{x} \equiv 2\pi^{-1/2}\!\int_{0}^{x}
\exp(-\eta^2) {\d}\eta\;$ is the error function \cite{Luke}. Then
$\CJ$ can be calculated by solving
\begin{equation}
  \label{CF}
  \Cf = 2\sqrt{\DA}\ierfs{(a_0 - \CA)/\CA}
      = 2\sqrt{\DB}\ierfs{(\CB-b_0)/\CB}
\end{equation}
and
\begin{equation}
  \label{CJ}
  \CJ = \CA\sqrt{\DA\over\pi}\exp\!\BRA{-\frac{\Cf^2}{4\DA}}
      = \CB\sqrt{\DB\over\pi}\exp\!\BRA{-\frac{\Cf^2}{4\DB}} \,,
\end{equation}
where $\CA$ and $\CB$ are some constants controlling the form of $\RA$
and $\RB$ outside the reaction zone; specifically, for $x \ll \xf - w$
we have
\begin{equation}
  \label{ra}
   \RA(x,t) = a_0 -  \CA\SQR{ \erf{x / \!\sqrt{4\DA t}} + 1}\,,
\end{equation}
and for $x \gg \xf + w$ there is
\begin{equation}
  \label{rb}
    \RB(x,t) =  b_0 + \CB \SQR{\erf{ x / \!\sqrt{4\DB t}} - 1}\,.
\end{equation}

It was confirmed by several methods, including the renormalisation group
analysis \cite{RG-Rapid,RG-Front}, numerical simulations \cite{J-E} and
heuristic arguments \cite{Koza}, that the values of the exponents
$\alpha$, $\beta$, $\GA$ and $\GB$ as well as the form of the
scaling functions $\SA$, $\SB$ and $\SR$ do not depend on $\DA$, $\DB$,
$a_0$ and $b_0$ if the values of these parameters are nonzero. However,
when one of the components is immobile (or `static'), the asymptotic
kinetics of the reaction front can change dramatically. For example, in
the mean-field approximation the width of the front converges to a
stationary value, and the reaction rate at $\xf$ decreases as
$t^{-1/2}$, which corresponds to $\alpha = 0$ and $\beta = \frac{1}{2}$
\cite{Hav95,J-E}. We have therefore two asymptotic universality classes:
one characteristic for the `dynamic' systems in which both components
diffuse, and the `quasistatic' one observed if one of the diffusion
constants is zero. Henceforth we shall assume $\DA>0$, so the two
asymptotic universality classes will be distinguished by determining
whether $\DB=0$ or $\DB>0$.

Although the peculiar kinetics of the systems with $\DB = 0$ was
noticed quite early, so far nearly all of the research has been
concentrated on the systems in which both of the diffusion constants
$\DA$ and $\DB$ are nonzero. Closer inspection of the powerful methods
developed in the last decade to investigate the reaction-diffusion
systems reveals that only the scaling ansatz, which is a purely
mathematical concept, can be trivially employed to investigate both
kinds of the systems. However, since even this relatively simple
approach has so far been carried out only for the `dynamic' problem
\cite{G-R,CDC91}, the theories of the two asymptotic universality
classes remain practically disconnected.

Several factors have lead to this situation. Theoretical results
\cite{Bstatic} derived for the case $\DB = 0$ utilized the basic
property of such systems, immobility of particles B, in a way that
cannot be extended for systems with $\DB > 0$. On the other hand, most
of the fundamental techniques developed for the case $\DB>0$ has been
based on the quasistationary approximation which requires that the ratio
$\JA/\JB$ of two opposite currents of particles A and B entering the
reaction zone should asymptotically go to 1. This condition cannot be
met by the systems with $\DB=0$, as in this case $\JB(t) \equiv 0$.

The aim of our paper is to unify our understanding of these two kinds of
the reaction-diffusion systems. The procedure we are using consists in
detailed examination of the case $\DB=0$ and comparison of the results
with those already derived for $\DB>0$. We study the systems with
$\DB=0$ by means of various methods and for arbitrary (positive) values
of $a_0$, $b_0$, and $\DA$. In particular we show the counterpart of the
quasistationary approximation (\ref{STAC}) which should be used if $\DB
= 0$. We argue that the form of these equations, as well as their
boundary conditions, determine the properties of the asymptotic
universality classes. In subsequent sections we study these properties
for the systems with $\DB=0$ using the heuristic theory of
Ref.~\cite{Koza}, the scaling ansatz, the mean-field approximation, and
numerical analysis. Although some aspects of the systems with $\DB=0$
turn out to be the same as of those with $\DB>0$, our analysis implies
that these two cases should always be considered separately.


\section{The limit $\DB \to 0$}
\label{LIMIT}
Consider equations (\ref{cfeq}) -- (\ref{rb}) with the values of $\DA$,
$a_0$ and $b_0$ fixed at some positive values, and $\DB$ going to 0. For
physical reasons we expect that as $\DB$ goes to 0, $\Cf$ converges to
some nonzero limiting value, and so the argument of $\Phi$ on the right
hand side of (\ref{cfeq}) diverges to infinity. We can therefore use an
asymptotic property of the error function, $x(1-\erf{x})\exp(x^2) \to
1/\sqrt{\pi}$ as $x\to\infty$ \cite{Luke}, and reduce (\ref{cfeq}) to
\begin{equation}
  \label{cfeq0}
   \Phi\!\left(
             \frac{-\Cf}{2\sqrt{\DA}}
       \right)
   =
        \frac{2 a_0 \sqrt{\DA}}{\sqrt{\pi}b_0 \Cf}
             \,.
\end{equation}
Since $\Phi(x)$ diminishes monotonically from $\infty$ to $0$ as $x$
grows from $-\infty$ to $\infty$, the above equation has a unique,
positive solution. Consequently, (\ref{CF}) and (\ref{CJ}) imply that
$\CJ$ and $\CA$ also converge to some positive values, but $\CB$ rapidly
diverges to infinity.

One can now use (\ref{CF}), (\ref{CJ}) and (\ref{cfeq0}) to arrive,
after some algebra, at an important relation valid only if $\DB = 0$
\begin{equation}
  \label{relcfcj}
  b_0 \Cf = 2\CJ\,.
\end{equation}
This equation has a very natural physical interpretation. On the one
hand the total number $M(t)$ of reactions occurring by time $t$ is
asymptotically equal
\begin{equation}
  \int_0^t\! \JA(\tau) {\d}\tau \,\sim\,
  \int_0^t\! \CJ/\sqrt{\tau}{\d}\tau \,\sim\,
  2\CJ\sqrt{t}\,.
\end{equation}
On the other hand, however, for $x \gg \xf + w$ we have $\RB \sim b_0$,
and for $x \ll \xf - w$ we expect $\RB \sim 0$. Neglecting terms of
order $b_0 w \ll b_0t^{1/2}$ we thus conclude that $M(t)$ can as well be
estimated by
\begin{equation}
 \int_{-\infty}^\infty [\RB(x,t)-\RB(x,0)] {\d} x \,\sim\,
  b_0 \xf(t)  \,\sim\,
  b_0 \Cf \sqrt{t} \,,
\end{equation}
which leads to (\ref{relcfcj}).

Our theory is consistent with the numerical simulations of Larralde {\em
et al} \cite{Bstatic}, who considered the one dimensional system (i.e.
for $d<\dc$) with $\DB=0$, $\DA = 1/2$ and $a_0/b_0 = 1$. They found
that, respectively, for $t = 500$, 1000 and 5000 the value of $\xf(t)$
was approximately equal to $11\pm 0.5$, $16\pm 0.5$ and $36\pm 0.5$, so
that $\xf/\sqrt{t} \approx$ $0.492\pm0.022$, $0.506\pm0.016$ and
$0.509\pm0.07$, respectively.  This is in excellent agreement with our
equation (\ref{cfeq0}), whose numerical solution reads $\Cf \approx
0.5060$. Below, in Section \ref{NUMS}, we will also verify our theory
using the mean-field approximation (i.e.~for $d > \dc$).


\section{General consequences of the scaling ansatz}
\label{SectionSA}

Let $\DB =0$ and $\DA$, $a_0$ and $b_0$ take on arbitrary (positive)
values. Assume that the asymptotic solutions of the G\'alfi and R\'acz
problem (\ref{GR}) in the long-time limit take on the scaling form
(\ref{SA}) -- (\ref{SR}) with $\alpha < 1/2$. Inserting (\ref{SA}) and
(\ref{SB}) into (\ref{GR}) and taking the limit $t\to\infty$ we find
that at any $x$ such that $|x-\xf|\ll t^{1/2}$ there is
\begin{equation}
  \label{prop1}
    \PT{\RA}      \,\propto\, t^{-\GA - \alpha - 1/2}
   \quad\mbox{and}\quad
    \DA\PXX{\RA}  \,\propto\, t^{-\GA -2\alpha}\,.
\end{equation}
Hence, because $\alpha < 1/2$, in the limit $t\to\infty$ the term
$\partial\RA/\partial t$ becomes negligibly small compared to
$\DA\partial^2\RA/\partial x^2$. This implies that for $|x-\xf|\ll
t^{1/2}$ the form of the scaling functions can be determined by solving
\begin{equation}
   \label{QSTAC}
     \begin{array}{rcl}
       \DA \PXX{\RA} &=&   R\,,\\[2ex]
       -\PT{\RB} &=&   R    \,.
     \end{array}
\end{equation}
The appropriate boundary conditions for these equations read
\begin{equation}
  \label{bcro}
    \begin{array}{rcl}
      \partial\RA/\partial x \to -\JA(t),\;  \RB \to 0
       &\;\mbox{as}\;& x \to -\infty\,,\\[1ex]
      \RA \to 0,\; \RB \to b_0 &\;\mbox{as}\;& x \to +\infty\,,
    \end{array}
\end{equation}
where $\JA(t) \propto t^{-1/2}$ \cite{Koza}. Equations (\ref{QSTAC}) are
the general counterparts of the quasistatic approximation (\ref{STAC})
if $\DA>0$ and $\DB=0$.

The boundary conditions (\ref{bcro}) determine the form of the boundary
conditions for $\SA$ and $\SB$ except for a constant multiplier. We will
take advantage of the fact that we are at liberty to multiply $\SA$ and
$\SB$ by arbitrary constants (which can be compensated for by
appropriate changes in $\EA$ and $\EB$) and assume that the boundary
conditions for $\SA$ and $\SB$ read
\begin{equation}
 \label{bcS}
 \begin{array}{rcl}
  \SA'(z) \to -1,\;  \SB(z) \to 0 &\;\mbox{as}\;& z\to -\infty\,,\\[1ex]
  \SA (z)\to 0,\; \SB(z) \to 1 &\;\mbox{as}\;& z \to +\infty\,,
 \end{array}
\end{equation}
 where the prime denotes the derivative
with respect to $z$.
Equations (\ref{bcS}) immediately imply that
\begin{eqnarray}
 \label{rex1}
  \GB &=& 0\,,\\[1ex]
 \label{eb}
   \EB &=& b_0\,.
\end{eqnarray}
The diffusive current $\JA$ of particles A for $-\sqrt{\DA t} \ll x \ll
\xf-w$ is asymptotically expected to be equal to $\CJ/\sqrt{t}$
\cite{Koza}. On the other hand, however, we can calculate it by
inserting (\ref{SA}) into $\JA = -\DA\partial\RA/\partial x$, which
leads to
\begin{eqnarray}
 \label{rex2}
  \GA + \alpha &=& 1/2\,,\\[1ex]
 \label{ea}
  \EA &=& \Cw \CJ / \DA\,,
\end{eqnarray}
where we denoted $\Cw \equiv \lim_{t\to\infty} w(t)/t^{\alpha}$.

Upon inserting  the scaling ansatz into the first of equations
(\ref{QSTAC}) we come to $\GA + 2\alpha = \beta$. Combining it with
(\ref{rex2}) we arrive at
\begin{equation}
  \label{rex3}
  \beta - \alpha = 1/2\,,
\end{equation}
We thus see that the scaling ansatz imposes on the values of the scaling
exponents three relations (\ref{rex1}), (\ref{rex2}) and (\ref{rex3}).
Only the first of them takes on a different form if $\DB >0$, as in that
case, by symmetry, $\GA = \GB$ \cite{CDC91}.

Equations (\ref{QSTAC}) imply that inside the reaction zone
\begin{equation}
   \label{QSTAC2}
       \DA \PXX{\RA} + \PT{\RB} =  0\,.
\end{equation}
Inserting into it the scaling ansatz (\ref{SA}) and (\ref{SB}), and
carrying out our `standard' limiting procedure ($t\to\infty$ at any
$x$ such that $|x-\xf(t)|\ll t^{1/2}$) we conclude, after some algebra
involving (\ref{relcfcj}), that $\SA$ and $\SB$ are related by a simple
formula $ \SA''(z) \,=\, \SB'(z)$. Upon integrating this equation and
using the boundary conditions (\ref{bcS}) to determine the integration
constant we finally come to
\begin{equation}
 \label{sasb}
  \SA'(z) = \SB(z) - 1\,.
\end{equation}
Note that for $\DB>0$ the scaling functions $\SA$ and $\SB$ satisfy an
entirely different relation $\SA(z) = \SB(-z)$, which comes from the
asymptotic symmetry of the system. Note also that relations
(\ref{QSTAC2}) and (\ref{sasb}) are independent of $R$ and,
consequently, of $d$.

As a quick application of (\ref{sasb}) consider a one dimensional system
with $\DB=0$. It is known \cite{Bstatic} that in this case $\SB =
\frac{1}{2}(1+\erf{z})$. So (\ref{sasb}) and (\ref{bcS}) lead to
\begin{equation}
 \SA(z) = \half \ierfc{z} \equiv
 \half \left[ z \erf{z} - z + \pi^{-1/2} \exp(-z^2) \right] \,.
\end{equation}


\section{The mean-field systems with $\DB = 0$}

\label{SMFA}

Consider a system governed by (\ref{GR}) with $\DA > 0$, $\DB = 0$ and
$R = k\RA\RB$.  This particular form of $R$ immediately implies
$\GA + \GB = \beta$. Using now (\ref{rex1}), (\ref{rex2}) and
(\ref{rex3}) we conclude that
\begin{equation}
 \alpha = 0,\quad 
 \beta  = 1/2,\quad
 \GA = 1/2,\quad \mbox{and}  \quad
 \GB = 0\,.
\end{equation}
These values are consistent with numerical simulations \cite{Hav95} and
heuristic ar\-gu\-ments \cite{J-E}.

It is easy to see that if length and time are measured in units of
$\lambda= (\DA/kb_0)^{1/2}$ and $\tau = 1/k a_0$, respectively, then the
solutions of (\ref{QSTAC}) reduce to those obtained for the particular
case $\DA=a_0=b_0=k=1$. Thus, in investigating the mean-field
reaction-diffusion systems, it is sufficient to examine in detail only a
system with some convenient values of the material parameters $\DA$,
$a_0$, $b_0$ and $k$. The solutions for arbitrary values of these
parameters can be then easily found by an appropriate choice of the
units. This property guarantees that any asymptotic length $l$ satisfies
$l(\DA,a_0,b_0,k) = \lambda l(1,1,1,1)$. In particular, the asymptotic
width of the reaction zone is given by
\begin{equation}
  \label{w}
  w = \W\sqrt{\frac{\DA}{k b_0}}\,,
\end{equation}
where $\W$ denotes the asymptotic width of the reaction zone in the
system with $\DA=a_0=b_0=k=1$. Numerical estimation of this parameter
yields $\W\approx 1.47$ (see the next section for more details).

Essentially the same line of reasoning was used in Ref.~\cite{Koza} to
show that in the mean-field system with $\DB>0$ the asymptotic width of
the reaction zone is given by $w = \WW(\DA \DB)^{1/3}(k
\CJ)^{-1/3}t^{1/6}$. However, in that paper it was incorrectly assumed
that $\WW~\equiv~1$. We estimated the correct value of $\WW$
numerically, obtaining $\WW\approx1.38$.

Inserting now the scaling ansatz into (\ref{QSTAC}) and using $R =
k\RA\RB$ we conclude that $\SA$ and $\SB$ satisfy
\begin{eqnarray}
 \label{eSA}
    \SA'' &=&  \W^2 \SA \SB\,,\\[1ex]
 \label{eSB}
    \SB' &=& \W^2 \SA \SB\,,
\end{eqnarray}

Upon inserting (\ref{sasb}) into (\ref{eSA}) we arrive at the nonlinear
differential equation for the mean-field scaling function $\SA$
\begin{equation}
  \label{saeq}
  \SA'' = \W^2 \SA(\SA'+1)\,.
\end{equation}
We can use it to estimate the behaviour of $\SA(z)$ and $\SR(z)$ for $z
\gg 1$. In this region we expect $\SA' \ll 1$ and $\SB \approx 1$, so
(\ref{saeq}) reduces to $\SA'' = \W^2 \SA$, which implies
\begin{equation}
  \label{tailA}
  \SR(z) \propto \SA(z) \propto \exp(-\W z) \,.
\end{equation}
We can also investigate the tail of particles B which forms for
$z\ll-1$. In this region we can assume $\SA(z) \sim -z$, so that
(\ref{eSB}) reduces to $\SB' = -\W^2 z\SB$, which leads to
\begin{equation}
  \label{tailB}
  \SB(z) \propto \exp(-\half(\W z)^2) \quad \mbox{and} \quad
  \SR(z) \propto |z|\exp(-\half(\W z)^2)\,.
\end{equation}
Thus, if $\DB = 0$, the mean-field form of the scaling function $\SR(z)$
is asymmetric, whereas for $\DB > 0$ this function is always symmetric,
which is most easily seen in the symmetric case $\DA = \DB$ and $a_0 =
b_0$.

We will now investigate some properties of the limit $\DB \to 0$ which
could not be analysed within the framework of the general theory
presented in Section~\ref{LIMIT}. As we already showed in our previous
paper \cite{Koza}, if $\DB>0$ then the mean-field density of particles B
at $\xf$ is asymptotically proportional to $\DB^{-2/3}t^{-1/3}$. As this
quantity has to be less than $b_0$, we conclude that $\DB^{-2/3}
t^{-1/3} < \mbox{const}_1$. Therefore the time $t^*$ at which the 
mean-field system
enters the long-time regime must satisfy
\begin{equation}
  \label{time}
   t^* > (\DB)^{-2}\cdot\mbox{const}_2\,.
\end{equation}
Only for times satisfying this relation can one use the quasistatic
approximation (\ref{STAC}). However, as $\DB\to 0$, the right hand side
of (\ref{time}) diverges to infinity. Consequently, as $\DB$ goes to 0,
$t^*$ diverges to infinity and in the limiting case $\DB = 0$ the
kinetics of the system can never be described with the quasistatic
approximation equations (\ref{STAC}). Although we derived these
conclusions only within the mean-field approximation,  it is reasonable
to expect that $t^* \to \infty$ as $\DB\to 0$ in any initially separated
reaction-diffusion system.

To summarize the differences between the two asymptotic universality
classes, in Table~\ref{Tab1} we list their main properties in the
mean-field approximation. The data for the case $\DB > 0$ come from
References \protect\cite{G-R}, \protect\cite{Koza} and
\protect\cite{Linear}.

\begin{table}[hbt]
  \caption{Comparison of the two asymptotic universality classes in the
   mean-field approximation.}
  \label{Tab1}
  \begin{center}
     \begin{tabular}{|c|c|c|}
     \hline                     &
       $\DA>0$, $\DB>0$ &
       $\DA>0$, $\DB=0$ \\
    \hline
     Diff. eqs. for $\SA$ and $\SB$ &
      \parbox{4cm}{
       \begin{center}
          $\SA'' = \SA \SB$\\
          $\SB'' = \SA \SB$
        \end{center}
      } &
      \parbox{4cm}{
       \begin{center}
        $\SA'' = \W^2 \SA \SB$\\
        $\SB'  = \W^2 \SA \SB$
       \end{center}
      } \\
      Diff. equation for $\SA$ &
        $\SA'' =  \SA(\SA + z)$&
        $\SA'' = \W^2 \SA(\SA'+1)$\\
      \parbox{4cm}{
         \begin{center}
           $\SR(z)$, $z\gg w$\\
           $\SR(z)$, $z\ll w$
         \end{center}
      } & $z^{3/4}\exp(-\frac{2}{3}z^{3/2})$ &
      \parbox{4cm}{
         \begin{center}
         $\exp(-\W z)$\\
         $z\exp(-\half(\W z)^2)$
         \end{center}
      }
      \\
      $w(t)$ &
        $\WW (\DA \DB /k \CJ)^{1/3}t^{1/6}$ &
        $\W \sqrt{\DA/kb_0}$ \\
      $\alpha$   & 1/6 & 0\\
      $\beta $   & 2/3 & 1/2\\
      $\GA$ & 1/3 & 1/2\\
      $\GB$ & 1/3 & 0\\
     \hline
    \end{tabular}
  \end{center}
\end{table}
\renewcommand{\arraystretch}{1}


\section{Numerical results}
\label{NUMS}

To check the theory presented in the previous sections for the case $\DB
= 0$ we solved numerically, using the finite-difference FTCS
(Forward Time Centred Space) 
method, partial differential equations (\ref{GR}) with the mean-field
reaction rate $R=k\RA\RB$. We present the data obtained for $a_0 = 0.1$,
$b_0=0.1$, $\DA = 0.1$, and $k=0.02$. Other values of these parameters
yielded similar results.

First of all we verified the theory presented in section \ref{LIMIT}. In
Fig.~1 we show the plot of $\RA$ and $\RB$ in the vicinity of $\xf$ at
$t=10^7$. The dotted line was computed from (\ref{ra}). It perfectly
matches the numerical solutions up to $x\sim 700$, i.e.~outside the
reaction layer. Actually, in the region $-1000 < x < 697$, the relative
error is less than $7\!\cdot\!10^{-4}$. Also, the value of $\Cf$
computed from (\ref{cfeq0}) is 0.2263, whereas its numerical estimation
$\xf/\sqrt{t}$ for $t = 10^7$ is 0.2253.

Next we investigated the scaling properties of the considered system. As
some aspects of this problem were already considered \cite{Hav95}, we
present briefly only those results which are relevant to our paper.
In Fig.~2 we show the log-log plot of $w(t)$.  We can see that it
initially grows as $\sqrt{t}$, which is a typical short time limit
behaviour \cite{Haim91}, but beginning from $t\approx10^3$ it quickly
converges to a constant value. This enabled us to estimate $\W\approx
1.47$.

Finally, in Fig.~3 we present the scaling plot of $\sqrt{t}R \propto
\SR$ as a function of $z = (x-\xf)/w$, where we used (\ref{w}) with $\W
= 1.47$. The plots for $t=10^6$ and $10^7$ are practically
indistinguishable. Note also two facts. First, for $z > 2$ the semilog
plot of $\sqrt{t}R$ is linear in $z$, in accordance with
(\ref{tailA}). Second, $\SR(z)$ is discontinuous at $z=-\xf/w$, which
reflects the fact that $\RB(x,t)$ is discontinuous at $x=0$.





\begin{figure}
  \epsfysize 3.2in
  \epsfbox[0 160 567 670]{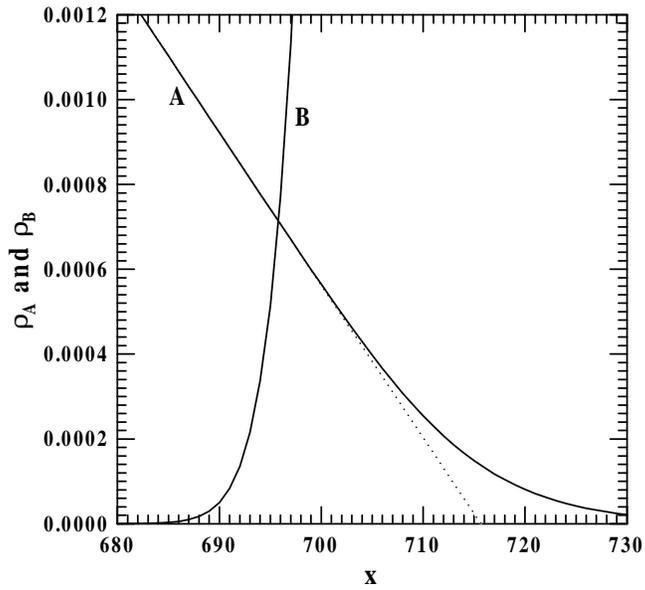}
  \caption{The
   concentrations $\RA$ and $\RB$ of A's and B's in the vicinity of
   $\xf\approx 713$ at $t=10^7$. The dotted line was computed from
   (\protect\ref{ra}).}
\end{figure}

\vfill 
\begin{figure}
  \epsfysize 3in
  \epsfbox[0 166 567 670]{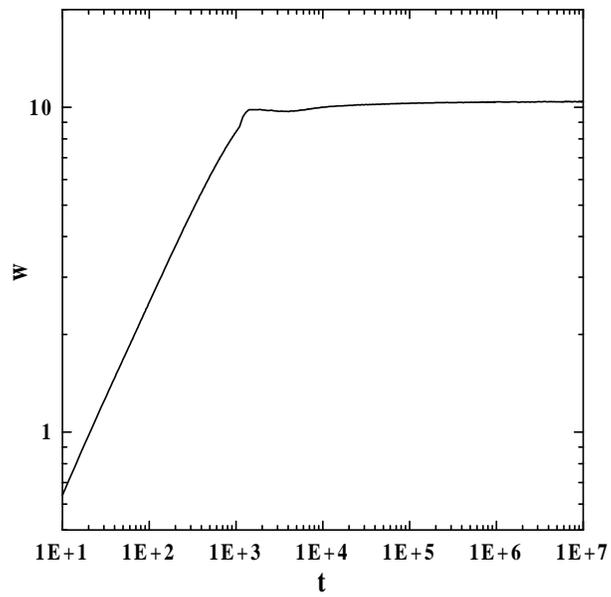}
  \caption{The log-log plot of the width of the reaction front $w$ as a
   function of time.}
\end{figure}

\clearpage

\begin{figure}[thp]
  \epsfysize 3.6in
  \epsfbox[0 166 567 670]{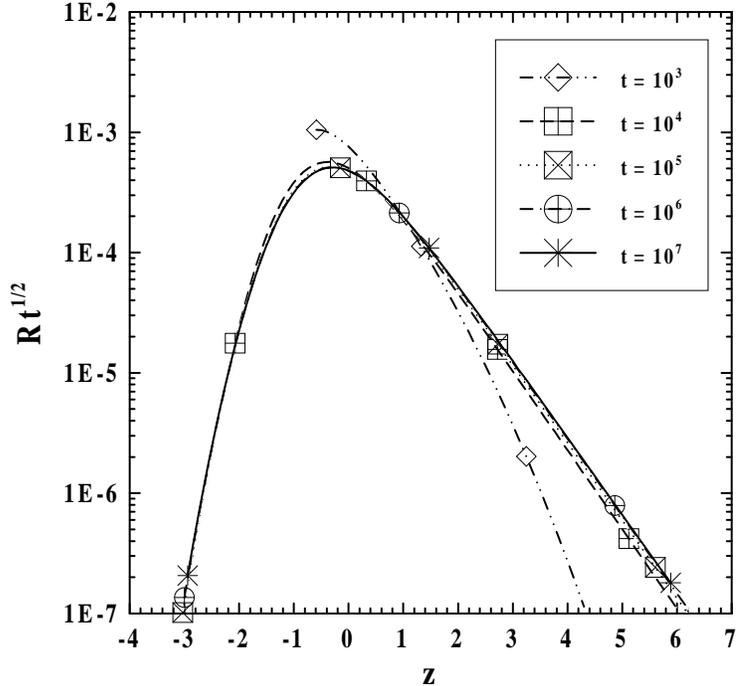}
  \caption{The scaling plot of $R\protect\sqrt{t}$ as a function of $z =
   (x-\xf)/w$.}
\end{figure}


\section{Summary and conclusions}

We have investigated the long-time behaviour of the concentrations $\RA$
and $\RB$ of species A and B in the initially separated diffusion
limited systems with $\DB = 0$. All of our analysis was carried out for
arbitrary values of $\DA$, $a_0$ and $b_0$.

First we derived the formulae (\ref{cfeq0}) and (\ref{relcfcj}) which
together with (\ref{ra}) describe the behaviour of $\RA(x,t)$ outside
the reaction zone. An interesting feature of these equations is that we
expect them to be valid for any reaction-diffusion system of the type
\ABOS\ that exhibits the scaling behaviour with $\alpha < \frac{1}{2}$.
Thus we conclude that many important properties of the
reaction-diffusion systems with $\DB=0$ depend only on the values of
$a_0$, $b_0$ and $\DA$, but not on the explicit form of $R$. This
includes the location of the reaction zone centre (controlled by $\Cf$),
the total reaction rate (controlled by $\CJ$), and the concentration
profile of particles A outside the reaction zone (controlled by $\CA$
and $\Cf$). A similar situation is observed in the systems with $\DB >
0$ \cite{Koza}.

Next we investigated general consequences of the scaling ansatz. We
concluded that it determines the value of $\GB = 0$ and imposes
relations (\ref{rex2}) and (\ref{rex3}) on the values of $\alpha$,
$\beta$ and $\GA$. These relations are also valid for $\DB>0$. We
proved that the scaling functions $\SA$ and $\SB$ are related by a
simple formula~(\ref{sasb}). We also concluded that for $\DB = 0$ the
asymptotic forms of $\RA$ and $\RB$ inside the reaction front can be
derived from equations (\ref{QSTAC}) which are to be solved with time
depending boundary conditions (\ref{bcro}). Thus these equations
determine the asymptotic properties of the reaction front.

We also examined in detail the properties of the reaction zone in the
mean-field approximation. In particular we determined the functional
forms of $\SA$, $\SB$ and $\SR$ far from $\xf$ and the dependence of the
width of the reaction zone on $k$, $a_0$, $b_0$ and $\DA$.

Our analysis showed that the main differences in the behaviour of the
initially separated reaction-diffusion systems with $\DB = 0$ and
$\DB>0$ arise from the fact that the term $\partial \RB/\partial t$
can be neglected in (\ref{GR}) only if the corresponding diffusion
constant $\DB$ is nonzero. Therefore, depending on whether  $\DB$ is
zero or not, the long-time behaviour of the reaction front is governed
by entirely different partial differential equations. If $\DB > 0$, then
we use the usual quasistationary equations (\ref{STAC}), otherwise we
must employ (\ref{QSTAC}). The different forms of these equations and
their boundary conditions imply the different forms of their solutions
and, consequently, different asymptotic properties of the two
universality classes. Therefore the cases $\DB=0$ and $\DB>0$ should
always be considered separately.

There is, however, evidence that in some special cases the two
asymptotic universality classes may  be very much alike. Such surprising
conclusion follows from the extensive numerical simulations of the one
dimensional system carried recently out by Cornell \cite{Cornell95}. He
considered a system with $\DA=\DB>0$ and concluded that $\alpha = 1/4$,
$\beta = 3/4$, and that asymptotically $R$ is a Gaussian centred at
$\xf$ with its width growing as $t^\alpha$. The same results were
derived analytically for the one dimensional system with $\DB=0$ by
Larralde {\em et al} \cite{Bstatic}. Note also that this form of $R$, or
more generally --- any $R$ that depends on $x$ and $t$ explicitly rather
than through $\RA(x,t)$ and $\RB(x,t)$, uniquely determines the form of
$\RA(x,t)$. This comes from the fact that whether or not $\DB = 0$,
asymptotical forms of $\RA$ and $R$ are related by the same differential
equation $\DA\partial^{2}\RA(x,t)/\partial x^{2} = R(x,t)$ with the same
boundary conditions. Therefore it is quite possible that the only
difference between the one dimensional systems with $\DB = 0$ and $\DB >
0$ lies in the form of $\SB$ and the value of $\GB$. It should be
stressed, however, that the asymptotic kinetics of one dimensional
systems with $\DB>0$ has recently become the subject of controversial
discussions \cite{Cornell95,LAHS3-8,Barkema,CommentC,CommentALHS}, and
further exploration of this topic is still required before the final
conclusions can be made.

\vspace{1cm}
\noindent

{\bf Acknowledgments} \nopagebreak \\ \nopagebreak
This work was supported by University of Wroc{\l}aw Grant No
2115/W/IFT/95.





\begin{thebibliography}{10}

  \bibitem{G-R} L. G\'alfi and Z. R\'acz, {\em Phys. Rev. A}
          {\bf 38}  (1988) 3151.

  \bibitem{Overview} Proceedings of the NIH Meeting on {\sf Models of
           Non-Classical Reaction Rates}, {\em J.~Stat. Phys.} {\bf 65},
           No. 5/6, 1991.

  \bibitem{HaimExperiment} H. Taitelbaum, Y. E. L. Koo, S. Havlin, R.
          Kopelman and G. H. Weiss, {\em Phys. Rev. A} {\bf 46} (1992)
          2151.

  \bibitem{Hav95} S. Havlin, M. Araujo, Y. Lereach, H. Larralde,
          A.~Shehter, H.~E.~Stanley, P.~Trunfio and B. Vilensky, {\em
          Physica A} {\bf 221} (1995) 1.

  \bibitem{Koza96} Z. Koza and H. Taitelbaum,  {\em Phys. Rev. E} {\bf
          54} (1996) R1040.

  \bibitem{Experiment} Y. E. Koo and R. Kopelman, {\em J. Stat. Phys.}
          {\bf 65} (1991) 893.

  \bibitem{HaimKudowa} H. Taitelbaum in {\em Diffusion Processes:
          Experiment, Theory and Simulations}, Andrzej P\c{e}kalski
          (Ed.), Springer-Verlag, Berlin, 1994.

  \bibitem{CD-Steady} S. Cornell and M. Droz, {\em Phys. Rev. Lett.}
          {\bf 70}   (1993) 3824.

  \bibitem{Krapivsky} P. L. Krapivsky, {\em Phys. Rev. E} {\bf 51}
          (1995) 4774.

  \bibitem{CDC91} S. Cornell, M. Droz and B. Chopard,
          {\em Phys. Rev. A} {\bf 44} (1991) 4826.

  \bibitem{RG-Rapid} B. P. Lee and J. Cardy, {\em Phys. Rev. E} {\bf 50} 
          (1994) R3287.

  \bibitem{RG-Front} M. Howard and J. Cardy, {\em J. Phys. A: Math.
           Gen.} {\bf 28}  (1995) 3599.

  \bibitem{RG-Homo} B. P. Lee and J. Cardy, {\em J. Stat. Phys.} {\bf 80}
          (1995) 971.

  \bibitem{Bstatic} H. Larralde, M. Araujo, S. Havlin and H. E.
           Stanley, {Phys. Rev. A} {\bf 46} (1992) R6121.

  \bibitem{Cornell95} S. J. Cornell, {\em Phys. Rev. E} {\bf 51} (1995)
           4055.

  \bibitem{LAHS3-8} M. Araujo, H. Larralde, S. Havlin and H. E. Stanley,
          {\em Phys. Rev. Lett.} {\bf 71} (1993) 3592.

  \bibitem{Barkema} G. T. Barkema, M. J. Howard and J. L. Cardy,
          {\em Phys. Rev. E} {\bf 53} (1996) R2017.

  \bibitem{Argentina} M. Hoyuelos, H. O. M\'{a}rtin and E. V. Albano,
          {\em J. Phys. A: Math. Gen.} {\bf 28} (1995) L483.

  \bibitem{BenRedner} E. Ben-Naim and S. Redner, {\em J. Phys. A:
          Math. Gen.} {\bf 28} (1992) L575.

  \bibitem{Koza} Z. Koza, {\em J. Stat. Phys.} {\bf 85} (1996) 179.

  \bibitem{Luke} Y. L. Luke, {\em Integrals of Bessel Functions},
          McGraw-Hill, New York, 1962.

  \bibitem{J-E} Z. Jiang and C. Ebner, {\em Phys. Rev. A} {\bf
          42} (1990) 7483.

  \bibitem{Linear} H. Larralde, M. Araujo, S. Havlin and H. E. Stanley,
          {\em Phys. Rev. A} {\bf 46} (1992) 855.

  \bibitem{Haim91} H. Taitelbaum, S. Havlin, J. E. Kiefer, B. Trus and
          G. H. Weiss, {\em J. Stat. Phys.} {\bf 65} (1991) 873.


  \bibitem{CommentC} S. J. Cornell, {\em Phys. Rev. Lett.} {\bf 75}
          (1995) 2250.

  \bibitem{CommentALHS} M. Araujo, H. Larralde, S. Havlin and H. E.
          Stanley, {\em Phys. Rev. Lett.} {\bf 75}  (1995) 2251.


\end{thebibliography}
\end{document}